%% file: tevcandidates.tex
\documentclass[a4paper,fleqn,usenatbib]{mnras}

\usepackage[T1]{fontenc}
\usepackage{ae,aecompl}

\usepackage{graphicx}	
\usepackage{amsmath}	
\usepackage{amssymb}	

\usepackage{times}

\usepackage{pdflscape}

\newcommand {\cms} {cm$^{-2}$ s$^{-1}~$}

\newcommand {\axg} {$\alpha_{\rm x \gamma}$ }
\newcommand {\awx} {$\alpha_{\rm wx}$ }

\newcommand {\ax} {\alpha_{\rm x}}
\newcommand {\ag} {\alpha_{\rm \gamma}}
\newcommand {\atev} {\alpha_{\rm VHE}}

\newcommand {\nufnu} {$\nu F_{\nu}$ }

\newcommand {\fermi} {{\it Fermi}-LAT }
\newcommand {\nh} {N$_{\rm H}$ }

\title[TeV-peaked candidate BL Lacs]{TeV-peaked candidate BL Lac objects}

\author[L. Costamante]{L. Costamante$^{1}$ \\
$^{1}$ASI -- Unit\`a Ricerca Scientifica, Via del Politecnico snc, I-00133, Roma, Italy 
}

\date{Accepted 2019 October 23. Received 2019 October 23; in original form 2019 August 23.}

\pubyear{2019}

\begin{document}
\label{firstpage}
\pagerange{\pageref{firstpage}--\pageref{lastpage}}
\maketitle

\begin{abstract}
BL Lac objects can be extreme in two ways:  with their synchrotron emission, peaking beyond 1 keV 
in their spectral energy distribution, 
or with their gamma-ray emission, peaking at multi-TeV energies up to and beyond 10-20 TeV, like 1ES\,0229+200.  
This second type of \emph{extreme BL Lacs} -- which we can name \emph{TeV-peaked BL Lacs} -- 
is not well explained by the usual synchrotron self-Compton scenarios for BL Lacs.
These sources are also important 
as probes for the intergalactic diffuse infrared background and cosmic magnetic fields,
as well as possible sites of production of ultra-high-energy cosmic rays and neutrinos.
However, all these studies are hindered by their still very limited number.
Here I propose a new, simple criterium to select the best candidates for TeV observations, 
specifically aimed at this peculiar type of BL Lac objects
by combining X-ray, gamma-ray and infrared data.
It is based on the observation of a clustering towards a 
high X-ray to GeV gamma-ray flux ratio, and it does not rely 
on the radio flux or X-ray spectrum.
This makes it suitable to find TeV-peaked sources also with very faint radio emission.
Taking advantage of the {\it Fermi} all-sky gamma-ray survey 
applied to the ROMA-BZCAT and Sedentary Survey samples,
I produce an initial list of 47 TeV-peaked candidates 
for observations with present and future air-Cherenkov telescopes. 

\end{abstract}




\section{Introduction}
BL Lac objects are a particular type of blazars, namely
radio-loud active galactic nuclei (AGN) with the relativistic jet pointing towards us.
They are the most copious emitters of very high energy radiation 
\citep[VHE, $\gtrsim$0.1 TeV, e.g.][]{vhe2008review} in the extragalactic sky,
especially after accounting for the absorption effect due to $\gamma$-$\gamma$ interactions 
with the diffuse Extragalactic Background Light 
\citep[EBL, see e.g.][and references therein]{hauserdwek,dominguez11,franceschini17}.
Their spectral energy distribution (SED) is dominated by non-thermal radiation, 
and is characterized by two distinctive broad humps, peaking at low and high energy, 
commonly (but not uniquely) explained as synchrotron and inverse Compton (IC) 
emission from a population of relativistic electrons in the jet.

BL Lac objects span a wide range of synchrotron peak frequencies, 
from infrared (IR) to X-ray energies, justifying the division in ``Low-" and ``High-frequency peaked" BL Lacs 
\citep[LBL and HBL, respectively,][]{padovanigiommi95},
or the more recent ``Low-", ``Intermediate-" and ``High-Synchrotron Peaked"  sources 
\citep[LSP, ISP and HSP,][]{abdoLSP}. 
Correspondingly, their high-energy peak in gamma-rays shifts from MeV up to TeV energies.  
This sequence of SEDs seems anti-correlated with the bolometric luminosity of the objects, 
forming the main part of the so-called ``blazar sequence" \citep{sequence2}.
Compton dominance plays a crucial role since its inception \citep[see][]{fossati98,finke13}.
However, 
it is still debated if this correlation has an underlying physical origin or it is just the result of selection biases, 
given that a significant fraction of the BL Lac population is still missing redshift measurements
\citep[see e.g.][for an alternative view]{simplified1}.

Extreme BL Lacs populate the highest-energy end of the blazar sequence. 
They come in two varieties, according to whether 
the extreme properties are shown in the low (synchrotron) or high energy emission.

The `BL Lacs extreme in synchrotron' are characterized by 
hard\footnote{Spectra are called hard or soft if the power-law spectral index
$\alpha<1$ or $\alpha>1$ respectively (i.e. photon index $\Gamma<2$ or $\Gamma>2$).
This convention corresponds to spectra rising or decaying with frequency in the SED.}
synchrotron X-ray spectra ($\ax<1$) at least up to $\sim$1 keV, 
locating the synchrotron peak energy above 1 keV, an order of magnitude higher than regular HBL.
In these objects the peak energy can reach 100 keV and beyond, as in 1ES\,1426+428. 
These objects are called ``Extreme synchrotron BL Lacs"  \citep{costamante01}  
or ``Extreme HSP"  \citep[EHSP,][]{arsioli18}. 
Some authors define EHSP the objects with $\nu_{peak}>10^{17}$ Hz (corresponding to 0.4 keV), but 
here we will adopt the original stricter definition of $>$1 keV, 
which has the advantage of a more direct assessment through the shape of the X-ray spectrum 
instead of a model of the SED.

The `BL Lacs extreme in gamma-rays' are instead characterized by 
hard intrinsic VHE spectra ($\atev<1$) up to and beyond $\sim$1 TeV, 
after correction for the EBL absorption effects. 
This occurs in some sources even considering the lowest possible EBL given by galaxy counts \citep[e.g.][]{natureEBL,0229hess},
locating their gamma-ray peak assuredly above 1 TeV, an order of magnitude higher than in regular HBL.
These sources are called ``{\it extreme-TeV BL Lacs}" \citep{tavecchio11} or ``{\it hard-TeV BL Lacs}" \citep{costamante17},
and have been introduced as a new class of BL Lac objects in \citet{bonnoli15}. 
They are also generically referred to as ``extreme HBL" (EHBL) 
although the two measures of extremeness --in synchrotron and gamma-rays-- do not always go 
together \citep[see e.g.][]{costamante17,foffano18}.
The prototype of this class is 1ES\,0229+200, which is the most extreme case discovered so far
with a hard VHE spectrum of $\atev\sim0.5$  and a gamma-ray peak well beyond 10 TeV.

While a synchrotron peak of even 100 keV or more represents no problem 
in standard shock-acceleration scenarios, 
being well below the maximum frequency that can be produced by electrons accelerated at the maximum possible rate
\citep[$h\nu_{peak}/\delta \sim1$-$10$ KeV $\ll 150$ MeV, $\delta$ being the beaming factor; see e.g.][]{felix_book},
a gamma-ray peak at multi-TeV energies is difficult to obtain by IC in blazars,
for standard one-zone leptonic models of the SED.
Both the decrease of the scattering efficiency in the Klein-Nishina regime 
and the lower energy density of the seed photons available for scatterings in the Thomson regime, 
as the gamma-ray energy increases, tend to steepen the TeV spectrum.
A synchrotron self-Compton (SSC) mechanism can still work,
but at the price of a narrow electron distribution which
does not reproduce the SED at UV and lower frequencies with a single population of electrons.
Furthermore, it requires  extremely high electron energies, 
very low radiative efficiency and conditions strongly out of equipartition by several orders of magnitude \citep{costamante17}.
Radically different alternative scenarios have been proposed,
from internal $\gamma$-$\gamma$ absorption on a Planckian
radiation  field \citep{felix08}  to a separate origin for the X-ray and TeV emissions,
the latter coming from kpc-scale jets \citep{bottcher08} or secondary emission from cascades \citep[e.g.][]{essey11,prosekin12}.

These extreme-TeV BL Lacs represent today a major challenge for the  
known acceleration and emission processes in  blazar's jets.
In addition, they are of great interest as probes of the near-infrared part of the 
diffuse EBL \citep[e.g.][]{costamante13,franceschini17}
and cosmic magnetic fields \citep[e.g.][]{dolag09,neronov10,finke15}, 
as well as possible sources of ultra-high-energy cosmic rays and neutrinos \citep[e.g.][]{padovani16,resconi17}.

However, their number is still limited to a handful, 
essentially because of the lack of systematic searches and of an all-sky survey at VHE with sufficient sensitivity.  
To identify more sources of this type among generic BL Lacs, the measurement of the VHE spectrum is mandatory, 
since it is the only way to determine the hardness or softness of their VHE emission
and with it the location of the gamma-ray peak. 
Sufficient photon statistics is also required to constrain the gamma-ray peak in a meaningful way. 
At present this is best achieved with imaging air Cherenkov telescopes arrays,
which however have the disadvantage of a narrow field of view and limited duty cycle, requiring dedicated campaigns. 

The purpose of this paper is to introduce a simple and handy criterium to select the most promising TeV-peaked candidates 
for pointed observations with present and future air-Cherenkov telescopes.
This is different from our previous work  \citep{costamante02} and from most of the recent selections 
\citep[e.g.][]{tavecchio10,massaro13,bonnoli15,chang17,arsioli18}.
The focus here is not on finding merely TeV-emitting sources, either normal or extreme synchrotron,
but rather \emph{TeV-peaked} sources,  namely those TeV-emitting BL Lacs with hard TeV spectrum 
and the gamma-ray peak above 1 TeV.

In the following, we refer to 
these objects equivalently as TeV-peaked BL Lacs or hard-TeV BL Lacs,  depending on whether 
the emphasis is on the physical SED properties or on the direct observational feature.
Unless otherwise stated, a flat $\Lambda$CDM cosmology is used with $h=\Omega_{\Lambda}=0.7$.
The VHE spectra are corrected for EBL absorption effects using the calculations by \citet{dominguez11}.
Spectral indices $\alpha$ are defined by flux density $S_{\nu}\propto \nu^{-\alpha}$ and correspondingly 
photon indices $\Gamma \equiv \alpha +1$ are defined by number density $N(E)\propto E^{-\Gamma}$.


\input{table_tevs.tex}

\section{Hard-TeV vs Soft-TeV BLLacs} 
According to the online catalog of TeV-detected sources (TeVCat\footnote{http://tevcat.uchicago.edu/}),
as of December 2018 there are 49 HBL detected at VHE. 
Out of these, 12 objects have no or uncertain redshift, and 4 objects do not have a published VHE spectrum yet.
It is thus not possible to derive their intrinsic EBL-corrected spectrum.
Of the remaining 33 HBL, 22 have soft TeV spectra while 8 show hard TeV spectra 
(namely with nominal $\Gamma_{\rm intr}\leq2$), though
two  might be considered borderline or transitional,
given the error on their slope and the almost flat spectrum (namely PKS\,0548-322 and H\,2356-309). 
The hard sources are listed in Table \ref{tevs}, along with their EBL-corrected spectral slope.

Based on this limited sample, hard-TeV BL Lacs seem to constitute about 1/4 of all detected HBL so far,
not considering for the moment 3 objects which have shown both hard and soft spectra in different datasets, 
namely Mkn 501 \citep{hegra501,magic501}, 1ES\,1727+502 \citep{1727magic,1727veritas} 
and  1ES\,1741+196 \citep{1741veritas,1741magic}.
These demonstrate however that a hard TeV spectrum is not always a permanent property,
and that HBL can undergo strong shift of the SED peak towards high energies
in the gamma-ray band as well as in the X-ray band. 

At high energies (HE, 0.1-100 GeV), 
the band where the {\it Fermi} Large Area Telescope \citep[LAT, $\sim$0.1--800 GeV,][]{lat}
is most sensitive,
all HBL display a hard spectrum (i.e. $\ag\lesssim 1$),
because the LAT is sampling the emission longward of the gamma-ray peak.
However, hard-TeV BL Lacs have generally a much lower flux in {\it Fermi}    
than soft-TeV objects, at similar SED luminosities.
As the gamma-ray peak shifts towards multi-TeV energies, the LAT band falls more and more inside the valley between 
synchrotron and Compton humps, where the emission is much weaker.
Regular soft-TeV HBL peak around 30--300 GeV, 
and therefore tend to be much brighter in {\it Fermi} for the same bolometric luminosity, 
because the peak is closer to the LAT passband.  Indeed, this was the case during the first years 
of {\it Fermi} operation: among the known TeV BL Lacs,  the first ones detected by \fermi were always characterized 
by soft intrinsic VHE spectra \citep[e.g.][]{abdo2009}.
As for any telescope, \fermi tends to detect sources with the peak emission in or close to its passband,
and for HBL this means a bias for the soft-TeV type.

To find new hard-TeV sources, therefore, somewhat counter-intuitively we need to look for objects with 
the \emph{lowest} --not the highest-- gamma-ray flux in \fermi, for a given synchrotron flux. 
The latter can be traced by the X-ray flux around 1 keV, which is close to the peak of the synchrotron emission in HBL.
The most promising TeV-peaked candidates, therefore, should be those with the largest X-ray to GeV flux ratio,
i.e. with the highest broad-band index \axg.  
In the following sections, 
this criterium is applied to a large sample of BL Lacs, 
and further refined with information from the Wide-field Infrared Survey Explorer (WISE).

\section{The sample and broad band indices}
The  sample of BL Lac objects considered here 
is constructed by merging the 5BZCAT blazar collection \citep{bzcat5} 
with the ``Sedentary Survey" sample, which selects by construction extreme HBL \citep{sedentary}.
Only established BL Lac objects with both X-ray and radio data  
are considered, for a total number of 1170 objects of every SED type.  
Nearly all of the 150 BLLacs in the Sedentary sample are comprised in 
the 5BZCAT, with the exception of three:  SHBL\,J101616.7+410812,
SHBL\,J142739.5-252102  and SHBL\,J224340.1-123100 (the latter detected by \fermi as 3FGL\,J2243.6-1230).
This total sample is then feeded with additional data taken from the gamma-ray (\fermi) and infrared (WISE) source catalogs.

The 1-100 GeV gamma-ray fluxes and photon indices are obtained from the Third \fermi AGN catalog \citep[3LAC,][]{3lac}.
For the objects not detected in the 3LAC, a reference flux of 8$\times 10^{-11}$ \cms 
is adopted,  with a photon index $\Gamma \equiv 1.7$ 
(appropriate for HBL, midway between the average index of the hard-TeV objects, $\sim$1.6, and soft-TeV ones, $\sim$1.8). 
This flux corresponds to a level just below 
the lowest flux detected in the 3LAC catalog (8.7$\times 10^{-11}$), and should be considered as an upper limit. 

The 0.1--2.4 keV X-ray fluxes are taken from the values listed in the catalogs with the following order of priority:  3LAC, Sedentary Survey, 5BZCAT.
Because the X-ray flux reported in the 5BZCAT is not corrected for galactic absorption, unlike 
the 3LAC and Sedentary Survey values, a multiplication factor is introduced to approximately compensate for the absorbed flux, 
as follows:  1.7$\times$ if the galactic column density \nh is $<5\times 10^{20}$ cm$^{-2}$, 2.0$\times$ if it is between 5 and 10$\times 10^{20}$,
and 3.0$\times$ if $\geq 10^{21}$ cm$^{-2}$. These values are derived from tests in XSPEC with photon indices 
in the range 1.5-2.3.
The galactic average \nh is derived from the \verb+nh+ ftool with the values from LAB survey maps \citep{kalberla}.

Monochromatic fluxes at 1 keV and 1 GeV are then calculated from integrated fluxes assuming a power-law spectrum,
using the spectral indices from the respective catalogs. 
For the X-ray spectral index, a value of $\alpha_{\rm X}=1.1$ is used, as in the Sedentary Survey, 
unless the gamma-ray photon index is $\Gamma>2.2$. In that case, $\alpha_{\rm X}=0.8$ is adopted, to better approximate
the slope of the X-ray emission in objects with LSP-type SED (which are characterized by steep GeV spectra). 

The infrared data are taken from the ALLWISE Source Catalog as provided by the NASA/IPAC Infrared Science Archive (IRSA),
with a cross-matching radius of 3.5\arcsec. 
The magnitude \verb+m3+ in the W3 filter measured with profile-fitting photometry (\verb+w3mpro+) is converted to flux density
at the effective frequency $\nu_{\rm iso}=2.6753\times 10^{13}$ Hz with the 
formula  $F_{\rm W3}(mJy)=31674 \times 10^{-m3/2.5}$ \citep{jarrett11}.
The infrared spectral index is calculated from the fluxes in the W3 and W2 filters.

The broad-band indices $\alpha_{\rm x \gamma}$ and $\alpha_{\rm wx}$ 
between the X-ray (1 keV), gamma-ray (1 GeV) and infrared (W3) frequencies
are then computed from the monochromatic fluxes according to the following formula:
\begin{equation}
\alpha_{12}\,\equiv \, - \frac{log\left( \frac{F_1}{F_2}*k \right)}{log\left(\frac{\nu_1}{\nu2}\right)}
\end{equation}
where $F_n$ is the flux at the frequency $\nu_n$ and $k=(1+z)^{\alpha_2 - \alpha_1}$ 
is the total K-correction factor, with $\alpha_1$ and $\alpha_2$ being the spectral indices at the respective frequencies.
The broad-band index is essentially the ratio of the monochromatic fluxes at two different energies, 
and represents the spectral index 
of a power-law spectrum connecting the two flux points \citep{stocke91,padovanigiommi95}.
For the K-correction when the redshift is unknown, 
a value of $z=0.3$ is assumed, which is the average redshift of HSP blazars in the 3LAC and Sedentary Survey samples.

\begin{figure*}
\centering
\includegraphics[width=15.0cm]{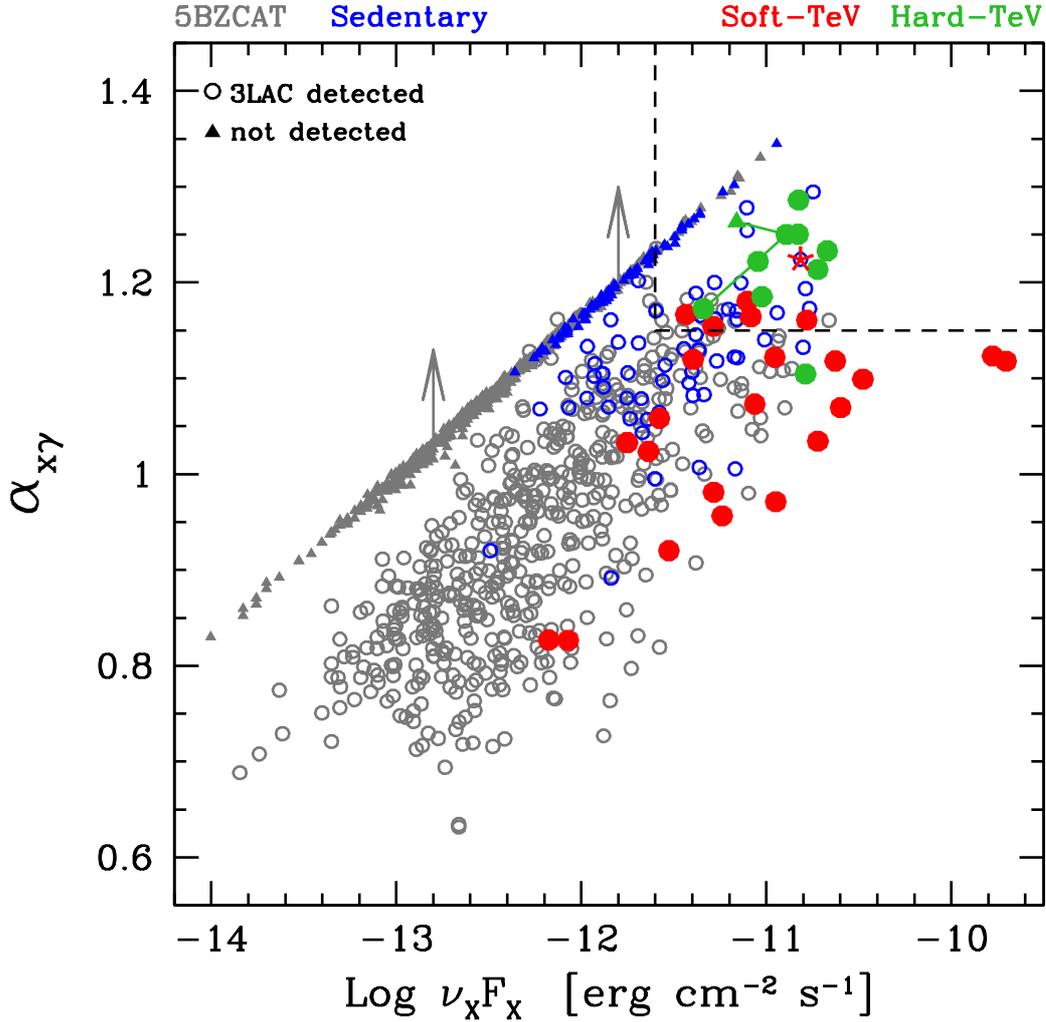}
\caption{
Plane of the broad-band index \axg vs the X-ray \nufnu flux at 1 KeV, 
for all BL Lacs in the Sedentary Survey and 5BZCAT samples (blue and grey markers, respectively). 
Gamma-ray data are obtained from the Fermi 3LAC catalog (open circles). For non-detected objects (triangles),
a reference flux of 8e-11 \cms  (in the 1-100 GeV band) is adopted as upper limit (see text). 
This translates to a lower limit to \axg, as indicated by the arrows.    
Given the correlation between the axes, the non-detected objects form a line in the figure 
(whose scatter is given by the K-correction).  
Hard-TeV BL Lacs (green full circles) cluster in the upper quadrant, while soft-TeV BLLacs (red full circles) show lower \axg values.
The red star marks the position of 1ES\,1426+428, which is extreme in synchrotron but not in TeV.
The green lines connect the positions of the prototypical hard-TeV BL Lac 1ES\,0229$+$200 
in 3 different states (Costamante et al. in preparation): 
1) during the first 2 years of {\it Fermi} operation (upper limit from the 2LAC catalog with same-epoch Swift data); 
2) detection in years 2011-2013 with average RXTE flux in the same epoch; 3) 3LAC catalog values as the other objects.
}
\label{fig1}
\end{figure*}

\section{TeV-peaked candidates selection}
Figure \ref{fig1} shows  \axg as a function of the X-ray flux, for all BL Lacs in the total sample.
Indeed \emph{the known hard-TeV BL Lacs cluster in the region of high \axg and high X-ray flux}.
We can interpret this property as follows:  
the index \axg works as a proxy for the location of the gamma-ray peak 
(namely, it tells how deep the LAT flux is inside the valley between the two SED humps),
while the X-ray flux works as a proxy for the synchrotron peak luminosity, 
in itself a tracer of the number of high-energy electrons in the source.
It thus gives an estimate  of the ``flux caliber" of the source for the detectability by Cherenkov telescopes.

Assuming a Compton dominance close to one (i.e. $L_{\rm IC}\sim L_{\rm synch}$),
sources with the higher \axg and X-ray flux
are more likely to have a gamma-ray peak at TeV energies and with higher flux.
The region of high \axg and high X-ray flux determined by the location of the known TeV-peaked objects 
can thus be considered a good criterium for the selection of new TeV-peaked candidates.  
To do so, a selection region is drawn at $\alpha_{\rm x \gamma}>1.15$ and Log($\nu_{\rm x} F_{\rm x}$)$> -11.6$ (see Fig. \ref{fig1}).
This box is drawn in a subjective way to include most of the known hard-TeV objects
and to obtain a sizeable but still manageable number of candidates.
Within this box, the probability of finding extreme TeV-peaked BL Lacs should increase to $\gtrsim 60$\%,
according to the ratio of Hard-TeV vs Soft-TeV sources detected so far.

However, a gamma-ray peak at multi-TeV energies is not the only reason for a BL Lac to have a large \axg.
Regular HBL 
can still be found inside the selection box
simply because of a lower Compton dominance,  due to -- for example -- higher magnetic fields in the emitting region.
In fact, considering a simple one-zone SSC scenario,  higher synchrotron peak frequencies would tend to
reduce the Compton dominance, all else being equal,  due to the decrease of seed photons available 
for Thomson scattering \citep[see for example Fig. 2 in][]{costamante02}.

In order to further reduce the chance of spurious objects, two additional cuts are introduced.
The first is based on the gamma-ray spectral slope, if the object is detected in the 3LAC or 
in the Third Catalog of Hard Fermi LAT Sources \citep[3FHL,][]{3fhl}:
namely, if the photon index $\Gamma_{\rm 3LAC}>2.2$ or $\Gamma_{\rm 3FHL}>2.4$. These values are larger than
the typical 1$\sigma$ error in these data, indicating that the source has most likely a steep GeV spectrum 
locating the gamma-ray peak below 1 and 10 GeV, respectively.
This cut excludes 6 objects.

\begin{figure}  
\centering
\includegraphics[width=9cm]{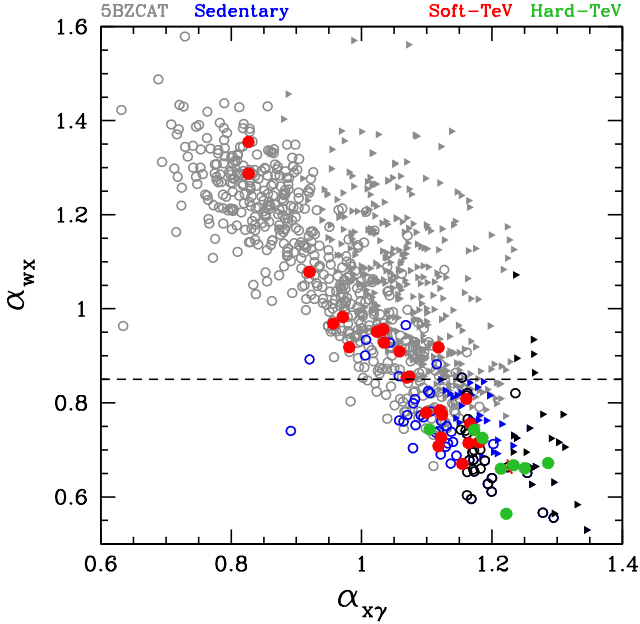}
\caption{
Colour-colour diagram of the broad-band spectral index \awx (between WISE W3
and 1 keV energies) vs \axg (between 1 KeV and 1 GeV). 
Labels and markers as in Fig. \ref{fig1}. Triangles represents lower limits 
to \axg for the objects not detected in the 3LAC. The dashed line shows the cut at \awx$<0.85$.
Black markers correspond to the objects inside the box in Fig. \ref{fig1}. 
}
\label{fig2}
\end{figure}

The second cut is based on the broad-band colour-colour diagram with the WISE infrared flux, plotted in Fig. \ref{fig2}.
The W3 filter is chosen as best compromise between infrared sensitivity and the minimization of the 
contribution of the host galaxy \citep[see e.g. Fig. 2 in][]{arsioli15}.   
The hard-TeV objects tend again to cluster towards \emph{low values} of \awx,  
indicating that they have a harder slope between infrared and X-ray energies than regular HBL. 
This is to be expected when the synchrotron emission peaks in the X-ray band, 
and the W3 flux is dominated by the non-thermal emission from the jet.
Nonetheless, the W3 band can still be contaminated by the host galaxy 
depending on the ratio between thermal and non-thermal emission in each object \citep{arsioli15,chang17}.
Since this contamination can keep \awx higher than it should,
the cut is applied rather loosely at $\alpha_{\rm wx}<0.85$ (see Fig. \ref{fig2}).
This cut reduces the number of candidates by a further 7 objects. 
None of these is detected in the 3LAC, while only two are marginally detected in the 3FHL 
with hints of steep spectrum, as expected.

The final list of 47 best candidates for a hard TeV emission is reported in Table 1, 
sorted according to their X-ray \nufnu flux 
and excluding those objects with published VHE spectrum.

\section{Comparison with previous selections}
Most criteria proposed so far 
for TeV observations
are developed to maximize the chance of detection, thus aiming at high VHE fluxes.
These criteria are less suited for finding hard-TeV objects, and in fact might be biased 
\emph{against} them.

Selections based on the \fermi  detection or spectrum  \citep[e.g.][]{tavecchio10}
are not very effective since all HBL tend to have hard LAT spectra, 
and a high \fermi flux  is more indicative of TeV softness rather than hardness, as shown before.
Selections based on SSC modeling  are generally flawed by construction, 
because 
the set of parameters and physical conditions valid for most HBL
are not able to reproduce the hard TeV spectra.
The shape of the SED in the synchrotron hump is very similar between normal and TeV-peaked HBL.
Without knowing a-priori that the source is TeV-peaked, 
the resulting TeV spectra are regularly steep,
especially when fitting the whole synchrotron hump from IR frequencies upwards \citep[see e.g.][]{costamante17}.
This type of SSC modeling provides therefore good predicting power  
only for soft-TeV HBL.

The curvature of the X-ray spectrum \citep{massaro11} does not seem a discriminating feature:
hard-TeV BL Lacs show curvature values near the peak in the range 0.2--0.4  \citep{costamante17}, 
similar to soft-TeV HBL.  

Selection schemes based on optical or IR colour diagrams alone, as the WISE blazar strip \citep{massaro13},
generally lose effectiveness for extreme BL Lacs \citep{chang17}.
As the SED peaks shift at higher energies, the jet luminosity in the optical/infrared bands
tends to become subdominant with respect to the thermal component of the host elliptical galaxy,
both because of the peak shift and because of the overall lower luminosity, if the blazar sequence holds.

The selection by \citet{bonnoli15} is based on a high X-ray to Radio flux ratio ($F_{\rm x}/F_{\rm r} >10^4$)
--similar to the value used for the Sedentary Survey and in \citet{costamante01}--
and an optical spectrum dominated by the host galaxy. 
The latter condition is achieved by using the list by \citet{plotkin}, which is also limited in redshift to $z\leq0.4$.
These criteria are effective to select extreme \emph{synchrotron} BL Lacs, 
but do not provide any direct indication of the location of the gamma-ray peak. 
Of the 9 objects in  \citet{bonnoli15}, three
are included in the present selection as well (see Table 1),
namely RBS\,0723, 1ES\,0927+500 and RBS\,0921. Two other objects are left out but by a small margin,
and thus can be considered good candidates also according to our criterium.
These are RBS\,1510 (out for \axg$=1.145$) and RBS\,1029 (out for $log(F_{\rm x})=-11.63$).

Recently \citet{foffano18} proposed a new selection.
It is focused again on extreme \emph{synchrotron} BL Lacs from the Swift-BAT 105-months source catalog
which are also detected by \fermi  in the 3LAC catalog.
The synchrotron peak position is estimated directly from archival Swift-XRT, Swift-BAT and BeppoSAX data.  
Their final sample of VHE targets have 8 objects in common with our total sample, 
5 of which are included in our selection as well.

\section{Conclusions}
The finding of TeV-peaked BL Lacs clustering towards high values of \axg and X-ray flux
and low values of \awx provides a simple criterium to select new TeV-peaked candidates. 
The broad-band index \axg is the most important parameter:
the limit on the X-ray flux is introduced to increase the chance of detectability at VHE 
and to reduce spurious sources, 
but can be relaxed for observations with more sensitive telescopes.

No filter on redshift has been introduced at this stage of the selection.
With the low EBL level close to galaxy counts confirmed by all recent analyses 
\citep[e.g.][]{natureEBL,dominguez11,costamante13,franceschini17,desai19},
detections at redshifts up to 1 and beyond are well possible even with the present generation of Cherenkov telescopes. 
However, with the observed fluxes and present telescopes, redshifts up to $z\sim$0.2-0.25  seem to provide 
the best compromise between space volume (i.e. number of objects) 
and flux around 1 TeV, allowing for spectral measurements extended into the TeV range.  
This is optimal for identifying hard-TeV sources, 
since a spectrum over at least a decade in energy can better constrain the slope 
of a power-law fit, thanks to the longer arm.
Furthermore, the EBL optical depth becomes less energy-dependent in the 1-8 TeV range, 
due to the shape of the EBL spectrum \citep{felix2001,costamante13}. 
This improves the chance of detection at the highest energies for intrinsically hard spectra.
Nevertheless, a higher source activity or more sensitive instruments 
like the future Cherenkov Telescope Array \citep[CTA,][]{cta} can allow comparable results at larger distances.

No VHE flux estimates are given at this stage. 
Even with the \fermi data, there is no clear way yet 
to distinguish between a hard-TeV and soft-TeV SED without the VHE data. 
The typical SSC modeling is thus of no use for predicting TeV-peaked sources, since  
it tends to systematically underestimate both flux and hardness of the spectrum in the TeV range.
Besides, in these objects SSC might not be the correct gamma-ray mechanism to begin with. 
 
An important advantage of the \axg criterium is to by-pass the radio information.
It is thus useful to explore also the faint end of the radio luminosity function,
where low-luminosity but still jetted AGN can be difficult to recognize in large radio and optical surveys.
Indeed, if the blazar sequence holds, the most extreme blazars might even be hidden in the radio quiet AGN population,
since the low radio flux from the jet might not be ``loud" enough 
with respect to the optical flux of the giant elliptical host galaxy \citep{ghisellini99,costamante07glast}.  

Both types of extreme blazars (synchrotron and TeV)
become brightest in the X-ray band. In this respect 
the results from the first all-sky survey in hard X-rays soon to be provided by eROSITA \citep{erosita12} 
will be optimal to find such objects, in particular when
coupled with the recently-released 8-years \fermi catalog \citep[4FGL,][]{4fgl},
the deepest all-sky survey so far in gamma-rays.
Still, TeV-peaked BL Lacs can be truly identified and studied only through VHE data.
Observations in the last decade have shown that essentially all BL Lacs (in fact all blazars) 
can copiously emit VHE gamma-rays, under proper conditions 
In this respect,  the importance of an unbiased extra-galactic sky survey at VHE at $\sim$1-2\% Crab-level sensitivity 
--as finally possible with CTA-- cannot be overstated, given also the new type of population studies 
(logN-logS, luminosity functions etc.)  and discovery potential that it opens.
Nonetheless, the present generation of Cherenkov telescopes retains a lot of potential for 
expanding and studying the population of TeV-peaked BL lacs.
Several other objects might exist, 
similar to 1ES\,0229+200 or 1ES\,1101-232 in flux and features, 
which simply have not been pointed yet.


%

\section*{Acknowledgements}
This research has made use of archival data, software and on-line services provided by the ASI Space Science Data Center (SSDC), 
and of data products from the Wide-field Infrared Survey Explorer, which is a joint project of the University of California, 
Los Angeles, and the Jet Propulsion Laboratory/California Institute of Technology, funded by the National Aeronautics and Space Administration.
This research has made use of the NASA/IPAC Extragalactic Database (NED), which is operated by the Jet Propulsion Laboratory, 
California Institute of Technology, under contract with the National Aeronautics and Space Administration.
This research has made use of data and software provided by the High Energy Astrophysics Science Archive Research Center (HEASARC), 
which is a service of the Astrophysics Science Division at NASA/GSFC and the High Energy Astrophysics Division 
of the Smithsonian Astrophysical Observatory.




\bibliographystyle{mnras}
\bibliography{biblio} 

\bsp	
\label{lastpage}


\input{hardtevs_table4land_A.tex}

\input{hardtevs_table4land_B.tex}


\appendix
\section{Data tables in ArXiv version}
This ArXiv version of the paper includes in the source files the 
following data, for better usability:
the ascii-only versions of Table 2,  Table 2 sorted by RA  
and an additional table obtained with no limit on the X-ray flux,
but with a stricter cut of $\alpha_{\rm x \gamma} \geq 1.2$,
sorted by decreasing X-ray flux.

\end{document}

%% file: table_tevs.tex
\begin{table}
\caption{Established Hard-TeV BL Lacs, as of December 2018.
The intrinsic photon index $\Gamma_{intr.}$ is the index of the power-law fit of the VHE spectrum 
corrected for EBL absorption according to \citet{dominguez11}, over the reported energy band.
The last column gives the references for the VHE data:
1)~\citet{0229hess}; 2)~\citet{0229veritas}; 3)~\citet{0347hess}
4)~\citet{0414hess}; 5)~\citet{0548hess}; 6)~\citet{0710veritas}; 7)~\citet{1101gerd}; 8)~\citet{natureEBL}; 9)~\citet{1218veritasLong};
10)~\citet{1218veritas}; 11)~\citet{2356hess2}; 12)~\citet{2356mio}. 
}
\label{tevs}
\centering
\begin{tabular}{l l c c c}
\hline
Name      &  $z$  &   $\Gamma_{intr.}$ &  Energy  & refs  \\
          &       &                    &     TeV      &          \\%
\hline 
1ES\,0229+200    & 0.140 &    $1.5\pm0.2$    &  0.6--12   & 1,2  \\
1ES\,0347$-$121  & 0.188 &    $1.8\pm0.2$    &  0.25--3   & 3   \\
1ES\,0414+009    & 0.287 &    $1.9\pm0.3$    &  0.15--2   & 4   \\  
PKS\,0548$-$322  & 0.069 &    $2.0\pm0.3$    &  0.3--4    & 5   \\
RGB\,J0710+591   & 0.125 &    $1.8\pm0.2$    &  0.3--4    & 6   \\
1ES\,1101$-$232  & 0.186 &    $1.7\pm0.2$    &  0.2--4 	  & 7,8   \\
1ES\,1218+304    & 0.182 &    $1.9\pm0.1$    &  0.2--4    & 9,10  \\ 
H\,2356$-$309    & 0.165 &    $1.95\pm0.2$   &  0.2--2    & 11,12 \\
\hline
\end{tabular}
\end{table}

%% file: hardtevs_table4land_A.tex
\begin{landscape}
\begin{table}
\caption{TeV-peaked BLLac candidates, sorted according to the X-ray flux considered as a tracer for their VHE flux. $F_{\rm X}$ is the RASS integral flux in the 0.1--2.4 keV band,
as listed in the catalogs. $\nu_{\gamma} F_{\gamma}$ and $\nu_{\rm X} F_{\rm X}$  are the derived monochromatic $\nu F_{\nu}$ fluxes at the respective energies.
The broad-band indices $\alpha_{\rm x \gamma}$ and $\alpha_{\rm wx}$ are calculated from the Wise (W3 filter), X-ray (1 keV) 
and gamma-ray (1 GeV) monochromatic fluxes, including the K-correction. The $\alpha_{\rm rx}$ index is copied from the 5BZCAT.
{\it Fermi}-LAT detection significances ($\sigma$) in the 3FHL and 3LAC catalogs
are also reported, together with the power-law photon index ($\Gamma$) from the 3LAC catalog. 
In the last column, positive values mark objects in the Sedentary Survey sample 
(2 if also the redshift is taken from the Sedentary sample because absent in the 5BZCAT).
}
\begin{tabular}{l c c c c c c c c c c c c }
\hline   
     Name &      Source Name      & $z$ & $F_{\rm [0.1-2.4~keV]}$ & $\nu_{\gamma} F_{\gamma}$ (1 GeV) & $\nu_{\rm X} F_{\rm X}$ (1 keV) & $\alpha_{\rm x \gamma}$ & $\alpha_{\rm rx}$ & $\alpha_{\rm wx}$ & $\sigma_{\rm 3FHL}$ & $\sigma_{\rm 3LAC}$ & $\Gamma_{\rm 3LAC}$  & Sedent.  \\
5BZCAT    &                       &     &  $erg\, cm^{-2} s^{-1}$   &  $erg\, cm^{-2} s^{-1}$         &  $erg\, cm^{-2} s^{-1}$        &                     &               &               &                     &                     &                      &  (a) \\
\hline
%
5BZBJ0123+3420 &          1ES 0120+340 &    0.272 &     6.14e-11 &   2.64e-13 &   1.79e-11 & 1.294 & 0.490 &   0.556 &       8.9 &       9.6 &    1.48 &         1     \\
5BZBJ0325-1646 &              RBS 0421 &    0.291 &     5.87e-11 &   1.45e-12 &   1.71e-11 & 1.173 & 0.460 &   0.654 &      16.3 &      18.9 &    1.79 &         1     \\
5BZBJ1031+5053 &          1ES 1028+511 &    0.361 &     5.58e-11 &   9.94e-13 &   1.63e-11 & 1.194 & 0.480 &   0.628 &      18.8 &      18.9 &    1.71 &         1     \\
5BZBJ0441+1504 & SHBL J044127.4+150456 &    0.109 &     3.93e-11 &   9.34e-14 &   1.15e-11 & 1.345 & 0.450 &   0.529 &       -   &       -   &     -   &         2     \\
5BZBJ0227+0202 &              RBS 0319 &    0.457 &     3.93e-11 &   1.10e-12 &   1.15e-11 & 1.168 & 0.500 &   0.596 &       6.9 &      11.1 &    2.04 &         1     \\
5BZGJ0643+4214 &           B3 0639+423 &    0.089 &     3.18e-11 &   9.34e-14 &   9.28e-12 & 1.330 & 0.580 &   0.584 &       -   &       -   &     -   &         0     \\
5BZBJ0509-0400 &          1ES 0507-040 &    0.304 &     2.70e-11 &   2.09e-13 &   7.88e-12 & 1.254 & 0.560 &   0.651 &       -   &       4.1 &    1.65 &         1     \\
5BZBJ1417+2543 &              RBS 1366 &    0.237 &     2.70e-11 &   6.56e-13 &   7.88e-12 & 1.181 & 0.570 &   0.717 &      10.1 &       6.5 &    2.16 &         1     \\
5BZBJ0930+4950 &          1ES 0927+500 &    0.187 &     2.69e-11 &   1.51e-13 &   7.85e-12 & 1.278 & 0.490 &   0.566 &       6.7 &       5.1 &    1.45 &         1     \\
5BZBJ1422+5801 &          1ES 1421+582 &    0.635 &     2.50e-11 &   4.45e-13 &   7.29e-12 & 1.200 & 0.470 &   0.611 &       7.8 &       7.0 &    2.02 &         1     \\
5BZGJ1504-0248 &      QUEST J1504-0248 &    0.217 &     2.46e-11 &   9.34e-14 &   7.18e-12 & 1.309 & 0.546 &   0.775 &       -   &       -   &     -   &         0     \\
5BZGJ0425-0833 &       EXO 0423.4-0840 &    0.039 &     2.40e-11 &   9.34e-14 &   7.00e-12 & 1.311 & 0.614 &   0.706 &       -   &       -   &     -   &         0     \\
5BZBJ1136+6737 &   RX J1136.5+6737$^a$ &    0.136 &     2.38e-11 &   7.11e-13 &   6.94e-12 & 1.161 & 0.540 &   0.734 &      14.5 &      15.3 &    1.72 &         1    \\
5BZBJ0847+1133 &          RBS 0723$^a$ &    0.198 &     2.38e-11 &   6.21e-13 &   6.94e-12 & 1.170 & 0.520 &   0.695 &      11.6 &      10.5 &    1.74 &         1    \\
5BZGJ2042+2426 &      RGB J2042.1+2426 &    0.104 &     2.32e-11 &   7.32e-13 &   6.77e-12 & 1.159 & 0.614 &   0.741 &       8.1 &       8.4 &    1.87 &         0     \\
5BZBJ0314+0619 &              RBS 0400 &    -     &     2.29e-11 &   9.34e-14 &   6.68e-12 & 1.301 & 0.520 &   0.717 &       5.9 &       -   &     -   &         1     \\
5BZBJ1456+5048 &              RBS 1444 &    0.479 &     2.21e-11 &   9.34e-14 &   6.45e-12 & 1.295 & 0.422 &   0.565 &       -   &       -   &     -   &         0     \\
5BZBJ2357-1718 &              RBS 2066 &    -     &     2.14e-11 &   5.35e-13 &   6.24e-12 & 1.172 & 0.550 &   -     &       7.0 &       8.6 &    1.80 &         1     \\
5BZBJ0837+1458 &  87GB 083437.4+150850 &    0.278 &     1.95e-11 &   9.34e-14 &   5.70e-12 & 1.291 & 0.572 &   0.724 &       -   &       -   &     -   &         0     \\
5BZBJ0506-5435 &               RBS 621 &    -     &     1.69e-11 &   6.13e-13 &   5.51e-12 & 1.152 & 0.556 &   0.743 &      15.2 &      14.9 &    1.60 &         0     \\
5BZBJ1535+5320 &          1ES 1533+535 &     0.89 &     1.79e-11 &   3.03e-13 &   5.22e-12 & 1.200 & 0.510 &   0.641 &       -   &       4.1 &    1.96 &         2     \\
5BZBJ0244-5819 &               RBS 351 &    0.265 &     1.57e-11 &   4.39e-13 &   4.58e-12 & 1.163 & 0.631 &   0.765 &      11.1 &      10.4 &    1.70 &         0     \\
5BZGJ1804+0042 &         RGB J1804+007 &     0.07 &     1.52e-11 &   9.34e-14 &   4.44e-12 & 1.277 & 0.695 &   0.776 &       -   &       -   &     -   &         0     \\
5BZBJ0110-1255 &               RBS 161 &    0.234 &     1.51e-11 &   4.34e-13 &   4.41e-12 & 1.165 & 0.510 &   0.678 &       -   &       5.8 &    1.93 &         1     \\
5BZBJ2343+3439 & SHBL J234333.8+344004 &    0.366 &     1.51e-11 &   3.70e-13 &   4.41e-12 & 1.171 & 0.550 &   0.679 &       8.9 &       6.8 &    1.75 &         1     \\
5BZGJ1056+0252 &              RBS 0921 &    0.236 &     1.49e-11 &   9.34e-14 &   4.35e-12 & 1.272 & 0.440 &   -     &       -   &       -   &     -   &         1     \\
5BZBJ1516-1523 &        RXSJ15163-1523 &    -     &     1.46e-11 &   9.34e-14 &   4.26e-12 & 1.269 & 0.470 &   -     &       -   &       -   &     -   &         1     \\
5BZBJ0018+2947 &              RBS 0042 &      0.1 &     1.43e-11 &   3.00e-13 &   4.17e-12 & 1.189 & 0.550 &   0.677 &       7.9 &       5.0 &    1.86 &         1     \\
5BZBJ0216+2314 &              RBS 0298 &    0.288 &     1.40e-11 &   9.34e-14 &   4.08e-12 & 1.266 & 0.560 &   0.665 &       6.1 &       -   &     -   &         1     \\
5BZBJ2138-2053 &              RBS 1769 &     0.29 &     1.33e-11 &   9.34e-14 &   3.88e-12 & 1.262 & 0.500 &   0.658 &       -   &       -   &     -   &         1     \\
5BZBJ0503-1115 & SHBL J050335.3-111507 &    -     &     1.29e-11 &   9.34e-14 &   3.76e-12 & 1.260 & 0.490 &   0.627 &       -   &       -   &     -   &         1     \\
5BZBJ1848+4245 &         RGB J1848+427 &    -     &     1.28e-11 &   2.67e-13 &   3.73e-12 & 1.183 & 0.574 &   0.660 &       6.4 &       6.3 &    1.66 &         0     \\
5BZBJ0014-5022 &              RBS 0032 &    -     &     1.23e-11 &   3.66e-13 &   3.59e-12 & 1.162 & 0.523 &   0.604 &       -   &       5.7 &    1.92 &         0     \\
5BZBJ0140-0758 &              RBS 0231 &    -     &     1.19e-11 &   9.34e-14 &   3.47e-12 & 1.254 & 0.550 &   0.664 &       -   &       -   &     -   &         1     \\
5BZBJ1257+2412 &          1ES 1255+244 &    0.141 &     1.17e-11 &   9.34e-14 &   3.41e-12 & 1.257 & 0.520 &   0.667 &       -   &       -   &     -   &         1     \\
5BZGJ0303+0554 &              RBS 0384 &    0.196 &     1.16e-11 &   9.34e-14 &   3.38e-12 & 1.255 & 0.560 &   0.750 &       -   &       -   &     -   &         1     \\
5BZBJ1021+1625 &        BZB J1021+1625 &    -     &     1.15e-11 &   9.34e-14 &   3.37e-12 & 1.252 & 0.540 &   0.695 &       -   &       -   &     -   &         0     \\
5BZBJ1008+4705 &         RXJ10081+4705 &    0.343 &     1.13e-11 &   9.34e-14 &   3.30e-12 & 1.249 & 0.460 &   -     &       -   &       -   &     -   &         1     \\
%
%
%
\hline
\multicolumn{13}{l}{$^a$ Announced detected at VHE, but not published yet.}
\end{tabular}
\end{table}
\end{landscape}

%% file: hardtevs_table4land_B.tex
\begin{landscape}
\begin{table}
\contcaption{}
\begin{tabular}{l c c c c c c c c c c c c }
\hline   
     Name &      Source Name      & $z$ & $F_{\rm [0.1-2.4~keV]}$ & $\nu_{\gamma} F_{\gamma}$ (1 GeV) & $\nu_{\rm X} F_{\rm X}$ (1 keV) & $\alpha_{\rm x \gamma}$ & $\alpha_{\rm rx}$ & $\alpha_{\rm wx}$ & $\sigma_{\rm 3FHL}$ & $\sigma_{\rm 3LAC}$ & $\Gamma_{\rm 3LAC}$  & Sedent.  \\
5BZCAT    &                       &     &  $erg\, cm^{-2} s^{-1}$   &  $erg\, cm^{-2} s^{-1}$         &  $erg\, cm^{-2} s^{-1}$        &                     &               &               &                     &                     &                      &  (a) \\
\hline
%
% ...
%
5BZBJ1008+4705 &         RXJ10081+4705 &    0.343 &     1.13e-11 &   9.34e-14 &   3.30e-12 & 1.249 & 0.460 &   -     &       -   &       -   &     -   &         1     \\
5BZBJ2123-1036 &         BZBJ2123-1036 &    -     &     1.08e-11 &   9.34e-14 &   3.15e-12 & 1.247 & 0.688 &   -     &       -   &       -   &     -   &         0     \\
5BZBJ1237+3020 &         RXJ12370+3020 &      0.7 &     1.06e-11 &   9.34e-14 &   3.09e-12 & 1.238 & 0.470 &   -     &       -   &       -   &     -   &         2     \\
5BZBJ0422+1950 &        MS 0419.3+1943 &    0.516 &     9.98e-12 &   9.34e-14 &   2.91e-12 & 1.237 & 0.500 &   0.726 &       -   &       -   &     -   &         1     \\
5BZBJ2249-1300 &        RXSJ22491-1300 &    -     &     9.73e-12 &   9.34e-14 &   2.84e-12 & 1.240 & 0.490 &   -     &       -   &       -   &     -   &         1     \\
5BZBJ1319+1405 &         RGB J1319+140 &    0.573 &     9.33e-12 &   2.67e-13 &   2.72e-12 & 1.161 & 0.634 &   0.816 &       -   &       4.3 &    1.87 &         0     \\
5BZBJ1536+0138 &        MS 1534.2+0148 &    0.311 &     8.82e-12 &   9.34e-14 &   2.57e-12 & 1.232 & 0.624 &   0.695 &       -   &       -   &     -   &         0     \\
5BZBJ1636-1248 &        BZB J1636-1248 &    0.246 &     8.74e-12 &   9.34e-14 &   2.55e-12 & 1.233 & 0.570 &   0.730 &       -   &       -   &     -   &         1     \\
5BZBJ1140+1528 &        BZB J1140+1528 &    0.244 &     8.73e-12 &   8.41e-14 &   2.55e-12 & 1.236 & 0.640 &   0.821 &       7.3 &       5.2 &    1.40 &         0     \\
5BZBJ1757+7033 &        MS 1757.7+7034 &    0.407 &     8.68e-12 &   2.11e-13 &   2.53e-12 & 1.170 & 0.520 &   0.747 &       5.6 &       5.5 &    1.71 &         1     \\
\hline
\end{tabular}
\end{table}
\end{landscape}